# Review Helpfulness Scores vs. Review Unhelpfulness Scores: Two Sides of the Same Coin or Different Coins?

Yinan Yu

University of Oklahoma

Dominik Gutt

Erasmus University Rotterdam, Rotterdam School of Management

Warut Khern-am-nuai

McGill University - Desautels Faculty of Management

**Abstract**

Evaluating the helpfulness of online reviews supports consumers who must sift through large volumes of online reviews. Online review platforms have increasingly adopted review evaluating systems, which let users evaluate whether reviews are helpful or not; in turn, these evaluations assist review readers and encourage review contributors. Although review helpfulness scores have been studied extensively in the literature, our knowledge regarding their counterpart, review unhelpfulness scores, is lacking. Addressing this gap in the literature is important because researchers and practitioners have assumed that unhelpfulness scores are driven by intrinsic review characteristics and that such scores are associated with low-quality reviews. This study validates this conventional wisdom by examining factors that influence unhelpfulness scores. We find that, unlike review helpfulness scores, unhelpfulness scores are generally not driven by intrinsic review characteristics, as almost none of them are statistically significant predictors of an unhelpfulness score. We also find that users who receive review unhelpfulness votes are more likely to cast unhelpfulness votes for other reviews. Finally, unhelpfulness voters engage much less with the platform than helpfulness voters do. In summary, our findings suggest that review unhelpfulness scores are not driven by intrinsic review characteristics. Therefore, helpfulness and unhelpfulness scores should not be considered as two sides of the same coin.

# 1. Introduction

Over the years, online review systems have been important information sources for consumers making purchasing decisions (e.g., [1, 2]). However, information overload—which prevents review readers from processing the sheer number of reviews available—has become increasingly common more recently, especially among large online review platforms [3]. To alleviate this issue, review platforms often leverage their review evaluation systems, which allows review readers to offer peer evaluations by casting votes for reviews they deem helpful or unhelpful. In turn, platforms then display helpfulness/unhelpfulness scores to help review readers find good reviews.

Review evaluation systems have been studied extensively in the literature. In particular, conventional wisdom in research and practice has treated review helpfulness scores and review unhelpfulness scores as opposites of each other. More specifically, review helpfulness scores are commonly used to identify high-quality reviews (e.g., [4, 5]), and review unhelpfulness scores are used to identify low-quality reviews (e.g., [6, 7]) or supplement review helpfulness scores to help readers determine review quality (e.g., [8, 9]). However, most previous studies in this stream of research have focused on understanding the factors that influence only review helpfulness scores (e.g., [10, 11]).There are limited studies that pay attention to the generation of review unhelpfulness scores. Because of this gap in the literature, the internal validity of using review helpfulness scores and review unhelpfulness scores to measure review quality remains unclear, despite the popularity of this approach. In this paper, we aim to address this research gap by using observational data obtained from a large restaurant review platform. Specifically, we propose the following overarching research question: *Is a review unhelpfulness score an opposite counterpart to a review helpfulness score?*

We investigate this research question in two ways. First, empirical studies have demonstrated that intrinsic review characteristics (e.g., review length, review extremity) are the primary factors that influence review helpfulness scores [10, 12]. However, even though there is scant evidence suggesting that intrinsic review characteristics (e.g., lack of information discussed in Connors, Mudambi and Schuff [13]) drive unhelpfulness scores, the Meta Goal-Based Choice Model [14] suggests otherwise. To resolve this tension, we propose our first research question:

*RQ1: Is a review unhelpfulness score influenced by intrinsic review characteristics that drive a review helpfulness score?*

We empirically answer this research question by examining the impact of two types of intrinsic review characteristics on a review helpfulness/unhelpfulness score. The first type is quantitative review measurements, which includes the review rating, length, and number of photos attached. The second type is textual features, which includes topic distribution, sentiment, and readability; we obtain all of these textual features using text mining techniques.

The second way we investigate the overarching research question involves users who cast review helpfulness/unhelpfulness votes. Although a few studies have examined the voting behavior that is associated with review helpfulness scores (e.g., [15]), there is virtually no empirical evidence on how unhelpfulness voters behave. Nevertheless, the psychology literature suggests that the underlying cognition usually differs between positive and negative evaluations [16]. In this regard, we propose our second research question as follows:

*RQ2: Are users who cast helpfulness votes similar to those who cast unhelpfulness votes?*

We answer this research question using two sets of analyses. First, we draw upon Balance Theory [17] to examine whether helpfulness and unhelpfulness voters are influenced by votes that they have received in the past. Next, we examine whether helpfulness voters and unhelpfulness voters behave similarly in terms of their respective platform involvement.

To operationalize our research agenda, we collaborate with a large restaurant review platform in Asia to obtain a dataset with information about review characteristics, review evaluations, and voter activities. Interestingly, our analyses demonstrate that review helpfulness scores are significantly different from review unhelpfulness scores. Unlike the helpfulness score, the review unhelpfulness score is not driven by most of the intrinsic review characteristics (e.g., only review rating exerts a U-shape impact on unhelpfulness scores in our study). Instead, unhelpfulness votes are driven by users who have received unhelpfulness votes themselves (i.e., the proportion of unhelpfulness votes that users received in the previous period has a significant positive impact on unhelpfulness votes that users sent in the subsequent period). Furthermore, unhelpfulness voters are much less engaged with the platform than helpfulness voters are (i.e., unhelpfulness voters tend to write fewer and shorter reviews, and also upload fewer photos). Our study is among the first to empirically examine how review unhelpfulness scores are generated differently from review helpfulness scores. Our results provide a theoretical foundation for future studies that seek to identify underlying mechanisms that drive review unhelpfulness scores, and for studies that caution against commonly adopted assumptions in the literature (e.g., review unhelpfulness scores are opposing counterparts of review helpfulness scores). Our findings also help platform managers understand that unhelpfulness scores do not consistently identify low-quality reviews. As such, using review unhelpfulness scores to identify reviews with low quality may in fact jeopardize a platform's content credibility.

# 2. Literature Review

In this section, we review the literature that is closely related to our study. In particular, we survey studies in the areas of review helpfulness scores, review evaluation, and review unhelpfulness scores.

## 2.1. Review Helpfulness Scores and Review Evaluation

Most papers in the research stream on review helpfulness scores have focused on the relationship between intrinsic review characteristics and review helpfulness scores. For example, a seminal paper by Mudambi and Schuff [10] demonstrates that review length and review valence significantly influence review helpfulness scores, whereas the product type (search goods vs. experience goods) moderates such influences. Later studies evaluate the impact of additional intrinsic review characteristics (e.g., ratings, readability, text sentiment, stylistic and semantic characteristics, embedded emotion) (e.g., [11, 18-23]).

Another related stream of literature has focused on predicting the helpfulness score of reviews based on review characteristics (as opposed to establishing a correlation or causal relationship between intrinsic review characteristics and review helpfulness scores, as we just discussed). Most studies in this substream of research have aimed to construct a numerical score that represents review helpfulness by using an algorithm that can automatically generate this score (as opposed to using user votes in a traditional review evaluation system). For instance, Kim, et al. [24] developed a model based on the Support Vector Machine (SVM) regression to predict the helpfulness score of reviews on Amazon; this score was based on structural, lexical, syntactic, and semantic characteristics. In the same vein, Xiong and Litman [25] showed that structural features of online reviews (e.g., review unigrams, metadata) are particularly useful in predicting review helpfulness scores.

Apart from studies that examine the relationship between intrinsic review characteristics and review helpfulness scores, another related substream of research has studied how external factors drive review helpfulness scores. These studies show that several reviewer characteristics (e.g., popularity, reviewer identification disclosure, engagement) significantly impact review helpfulness scores. For example, Kuan, et al. [19] found that reviews generated by reviewers with higher credibility—top reviewers recognized by a review platform—are more likely to obtain high helpfulness scores. Similarly, Forman, Ghose and Wiesenfeld [26] used Amazon review data to show that reviewers' disclosure of identity significantly affects helpfulness scores that their reviews receive. Other studies have found that additional reviewer characteristics, such as social interactions among reviewers and review readers, can impact review helpfulness scores [15].

As the literature has established strong connections between review helpfulness scores and several intrinsic review characteristics, review helpfulness scores have become the de facto measures of review quality in the literature. For example, several papers in multiple disciplines use review helpfulness scores to measure review quality, which is usually defined as the level of information diagnosticity that facilitates review readers' actions, such as purchasing decisions. [4, 5, 27-29]. In the next subsection, we review the literature for the counterpart of review helpfulness scores: review unhelpfulness scores.

## 2.2. Review Unhelpfulness Scores

Although researchers have studied review helpfulness scores extensively, review unhelpfulness scores have received far less attention. Notably, several studies have observed that review unhelpfulness scores help facilitate online review platforms' operations, such as review spam detecting [30] and review ranking [31]. Meanwhile, in one of the few studies that investigate the generation of review unhelpfulness scores, Connors, Mudambi and Schuff [13] use a controlled laboratory experiment to show factors that are correlated with review unhelpfulness scores; some of these factors are related to intrinsic review characteristics (e.g., "lack of information," "poor writing style"), whereas others are not as apparent (e.g., "low credibility"). Aside from these findings, however, little is known about review unhelpfulness scores' characteristics and capability, including whether review unhelpfulness scores are correlated with intrinsic review characteristics (and hence review quality) and, in such cases, significantly impact recipients' behavior [32].

Despite this knowledge gap, review unhelpfulness scores are surprisingly used widely to measure two aspects of review quality in the literature. For example, the computer science literature commonly uses review unhelpfulness scores as low-quality indicators to predict review helpfulness scores (e.g., [6, 7, 33]). Also, and more predominantly, many studies use scores of review unhelpfulness in conjunction with scores of review helpfulness to measure review quality (see Table A1 in the Appendix for further detail). Both uses of review unhelpfulness scores critically assume that review helpfulness scores and review unhelpfulness scores are not only cast similarly, but that they also represent intrinsic review characteristics (e.g., high-quality reviews have high helpfulness scores, low-quality reviews have high unhelpfulness scores). This paper explores research questions that can help validate these assumptions.

In summary, the literature has extensively studied review evaluation and review helpfulness scores, demonstrating that review helpfulness scores tend to relate directly to reviews' intrinsic characteristics. As a result, such scores could serve as consistent estimators of review diagnosticity (and hence review quality). Review unhelpfulness scores, in contrast, have been understudied, but nonetheless have been used in studies to indicate lower quality reviews. To validate this conventional wisdom, we useconduct empirical analyses on data from a restaurant review platform. In the next section, we discuss the theoretical background for our study and also our research hypotheses.

## 3. Theoretical Background and Research Hypotheses

In this section, we draw on established theories to develop a theoretical background that guides our empirical analyses. To this end, we combine a meta goal-based choice model [14] with balance theory [17] in a two-step theorization. The meta goal-based choice model summarizes a set of choice heuristics that humans apply due to limited mental processing capacity [13]. In particular, we use this model to help us answer our first research question: Are review helpfulness scores and review unhelpfulness scores driven by a similar set of intrinsic review characteristics? We then employ balance theory, which describes the actions that humans take to return to a balanced state of mind after an external factor has unbalanced it. We use this theory to address our second research question and investigate the discrepancy between helpfulness and unhelpfulness voters.

The literature has clearly established information cues that make reviews helpful. These cues include readability, length, sentiment [34], photos [35], and review content [18]. Both the presence of these cues themselves (e.g., photos) as well as their extent (e.g., long reviews) are perceived as helpful. In turn, the literature has also adopted a somewhat naïve assumption—that the absence of these cues (e.g., no photos, short reviews) would be unhelpful [6-9]. However, if helpful and unhelpful reviews are disjunct sets to a review reader, then voting for a review as helpful makes voting for another review as unhelpful superfluous: it would simply suffice to not vote helpful. Yet the question remains: how does a review reader decide whether to vote helpful or unhelpful? We draw upon a meta goal-based choice model to theoretically explain such a decision. According to this model, review readers presented with abundant online reviews face a significant cognitive burden: deciding whether to cast helpful or unhelpful votes for reviews. Presented with hundreds of reviews for a focal restaurant, should reviewers give helpful votes to all reviews that are rich in helpful cues and give all other reviews unhelpful votes? The meta goal-based choice model [14] suggests that in such situations, individuals resolve this problem by applying a choice heuristic guided by four principles: (1) maximizing choice accuracy, (2) minimizing cognitive effort required to make a choice, (3) minimizing negative emotion when making a choice, and (4) maximizing ease of justification for the choice. Bettman et al. [14] note that the relative weight of these principles for making decisions depends on the characteristics of the decision problem; that is, the four principles are not equally important to all decisions. The decision problem we study is primarily characterized by significant cognitive burden: reviewers face a very large number of reviews and must choose to cast either helpful or unhelpful votes, conditional on the choice to vote at all. Based on these characteristics, higher weight is placed on principle 2 (minimizing cognitive effort to make a choice) and principle 3 (minimizing negative emotion when making a choice).[1] With respect to principle 2, review readers are likely to opt for

---

[1] The characteristics of our decision problem put less weight on principles of maximizing justification for choices and maximizing choice accuracy. First, voters in our context do not need to justify their decisions to others because others cannot observe whether voters have in fact voted on a particular review; hence, they cannot be held

either of two options, conditional on casting a vote: vote information-rich reviews as helpful or vote information-poor reviews as unhelpful, but not both. Assuming disjunct sets of helpful and unhelpful reviews, indicating one of the options for a subset of reviews automatically classifies the remaining reviews into the other subset. With respect to principle 3 [36], review readers are more likely to vote for helpful reviews as helpful rather than vote for unhelpful reviews as unhelpful. In our context, this means that helpfulness voting should be much more prevalent than unhelpfulness voting, and that factors driving helpfulness scores are unlikely to symmetrically drive unhelpfulness scores in the opposite direction. Based on this reasoning, we formulate our first research hypothesis:

*H1: Factors that drive helpfulness scores do not drive unhelpfulness scores in the opposite direction.*

Although the meta goal-based choice model suggests an asymmetrical generation of review helpfulness scores and review unhelpfulness scores, the model does not explain how review unhelpfulness scores are generated. To obtain that explanation, we turn to balance theory. Conceptually, unhelpfulness voting represents a negative evaluation of user-generated content. Such negative experiences tend to unbalance consumers' states of mind [37]. Balance theory suggests that individuals will strive to restore equilibrium after their originally balanced state has become unbalanced [17], and research has found that balance theory contributes to user-generated content [38, 39]. Specifically, individuals may choose to return to a balanced state of mind by giving a negative evaluation to others after they receive a negative evaluation

---

accountable. Even though one could argue that voters are also accountable towards themselves, this should increase the weight of this principle. Second, due to missing accountability, there is very low need for voters to ensure maximization of voter accuracy, which is consistent with evidence from the literature [36] themselves. In our context, this means that when reviewers receive negative feedback, they are more inclined to give negative feedback to other reviews, which can explain the generation of review unhelpfulness scores. Based on this reasoning, we formulate our second research hypothesis as follows:

*H2: When reviewers receive unhelpfulness votes, they will also cast more unhelpfulness votes.*

Next, we present the context and data that we use to perform empirical analyses that are guided by the underpinning theories we have discussed in this section.

## 4. Research Context and Data

### 4.1 Research Context

We collaborated with a large restaurant review platform in Asia to investigate our research questions. Users of this platform contributed textual reviews, user-generated photos, and star ratings (ranging from 1 to 5) for their evaluation of the restaurant. Similar to other review platforms, this website allowed review readers to vote for each review as either "helpful" or "not helpful."

The platform provided us with three datasets. The first dataset contains information on reviews generated in 2016 (including review ID, reviewer ID, review date, textual content, associated star rating, and number of photos attached). The second dataset consists of the timestamps of helpfulness votes and unhelpfulness votes for each review in the first dataset. Although this dataset allowed us to track which review reader cast a helpfulness/unhelpfulness vote for a certain review at a specific time, the identity of voters was hidden from reviewers (i.e., reviewers could only observe the helpfulness/unhelpfulness scores that their reviews received and not the identity of the voters themselves). The third dataset consists of review reading logs (the review, the reviewer reader, and the date and time of reading) and allows us to control for the impact of exposure to the reviews, since reviews that have been read more often would naturally obtain more votes.

## 4.2 Data

Our primary dependent variables are *Helpfulness* and *Unhelpfulness,* which are the number of helpfulness/unhelpfulness votes attained by the focal review within 30 days of posting, respectively. In our first dataset, 157,285 reviews were generated on the platform in 2016.

We also include three sets of explanatory variables. First, we incorporate quantitative review characteristics. Conceptually, we attempt to examine the impact of review extremity, review depth, and review richness on votes received; to do so, we following the literature and measure them by star rating and quadratic term, number of characters, and number of photos, respectively [4, 10]. We also control for the number of times readers accessed the review in the first 30 days (*View*) and the chronological order of the review for a restaurant (*Rank*), so we may capture the potential impacts caused by exposure and review rank.

Second, we explore the content of review text to better understand how textual features affect helpfulness/unhelpfulness scores. Specifically, we conduct topic modeling, a sentiment analysis, and a readability analysis to examine the influence of review content. Our topic modeling is based on the Latent Dirichlet Allocation (LDA) Model [40], which allowed us to discover semantic structures hidden in the textual content of reviews in our sample. LDA is an unsupervised clustering model used to discover abstract topics in a collection of documents and generate a predefined number of topics. In LDA, each review is modeled as a mixture of various topics, which, in turn, are modeled as term distributions. We use the *scikit-learn* package for Python [41] to analyze our review text. We run the model with three, four, five, and six topics, respectively, and then inspect the distribution of terms. We find that the choice of four topics delivered the lowest perplexity, which commonly measures the evaluation of topic models [40], and delivered the most meaningful topic distribution. Table 1 lists the top ten terms of the four topics we used: food and meal, drink and dessert, service, and infrastructure.

**Table 1. Top Terms for Topics in Reviews**

| Topic | Top Terms |
|---|---|
| 1 Food and meal | Pork, fried, rice, delicious, eat, chicken, taste, noodle, sauce, soup |
| 2 Drink and dessert | Like, ice, eat, cream, sweet, good, tea, coffee, delicious, milk |
| 3 Service | Eat, food, good, order, like, come, time, restaurant, price, menu |
| 4 Infrastructure | Shop, park, restaurant, eat, come, road, open, locate, lot, store |

We next conduct a sentiment analysis to determine whether the polarity (i.e., whether a review is positive, negative, or neutral) and arousal (i.e., the level of energy characterizing an emotional experience) of review text affects helpfulness/unhelpfulness scores. We use TextBlob, a lexicon-based sentiment analysis tool in Python, to calculate a sentiment score for each review based on content [42]. The score, which is an average polarity of each word in that review, is in the range of –1 to 1, for which 1 represents a completely positive statement and –1 represents a completely negative statement. We measure arousal of review text by using a popular arousal dictionary developed by Warriner et al. [43], which is widely used in text mining (e.g., [44]). The dictionary contains 13,915 words with arousal ratings ranging from 1.6 to 7.79. We calculate each review's arousal rating by averaging the arousal rating of each word in that review.

Finally, we examine whether text readability affects review helpfulness/unhelpfulness scores. We use the Python package NLTK [45] to calculate the Gunning-Fog (GF) index, which estimates the years of formal education a person needs to understand the text on the first reading.

GF is commonly used in the IS literature to measure readability (e.g., [46]). A higher GF index indicates that a review is more difficult to understand.

Studies demonstrate that characteristics of reviewers, the information source, also impact the evaluation of review helpfulness (e.g., [47]). Accordingly, we control for the impact of the total number (*ReviewerReviewCnt*), average rating (*ReviewerRating*), and variance of ratings (*ReviewerVariance*) of reviews generated by the focal reviewer prior to the focal review. Users' perception of the focal restaurant, positive or negative, could spill over to their evaluation of reviews of that restaurant. We therefore incorporate a set of restaurant characteristics to control for such influence, including the total number of reviews (*RestReviewCnt*), average star rating (*RestRating*), and rating variance (*RestVariance*) for reviews generated for the restaurant prior to the focal review.

In Table 2, we report the summary statistics of all variables in our empirical analyses, including quantitative review characteristics, variables related to textual content of reviews, and variables that represent characteristics of the corresponding restaurant in the reviews. The average of helpfulness votes and unhelpfulness votes are about 5.98 and 0.02, with a large standard deviation of 11.01 and 0.14, respectively, which suggests that our sample contains reviews with sufficient variation in the number of votes received. Reviews in our sample obtained a rating of 3.87 stars on average, with roughly 505 characters in each review. On average, reviews came with five photos, had six reads, and ranked 39$^{th}$ in the list of a restaurant's reviews. Meanwhile, 24% of reviews' content, on average, focused on the restaurant's foods, 27% focused on drink and dessert, 31% focused on service, and the other 19% focused on the restaurant's infrastructure. The mean sentiment of the review text was 0.24, which is slightly positive. The mean values of text arousal and review readability scores were 4.06 and 8.52, respectively. We report correlations among these variables in Table A2 in our Appendix.

**Table 2. Summary Statistics**

| Variable | Mean | Std. Dev. | Min | Max |
|---|---|---|---|---|
| Dependent Variables | | | | |
| *Helpfulness* | 5.98 | 11.01 | 0 | 130 |
| *Unhelpfulness* | 0.02 | 0.14 | 0 | 3 |
| Independent Variables | | | | |
| *Length* | 504.50 | 575.55 | 2 | 23,257 |
| *Photo* | 5.39 | 5.82 | 0 | 128 |
| *% of Topic 1* | 0.24 | 0.28 | 0 | 1 |
| *% of Topic 2* | 0.27 | 0.34 | 0 | 1 |
| *% of Topic 3* | 0.31 | 0.31 | 0 | 1 |
| *% of Topic 4* | 0.19 | 0.24 | 0 | 1 |
| *Sentiment* | 0.24 | 0.18 | -1 | 1 |
| *Arousal* | 4.06 | 0.20 | 2.05 | 7.14 |
| *Readability* | 8.54 | 3.30 | 0.40 | 111.75 |
| *Rating* | 3.87 | 0.89 | 1 | 5 |
| *Rank* | 38.55 | 60.66 | 2 | 587 |
| Control Variables | | | | |
| *View* | 6.34 | 13.54 | 0 | 771 |
| *ReviewerReviewCnt* | 232.43 | 400.60 | 1 | 2,950 |
| *ReviewerRating* | 3.83 | 0.48 | 1 | 5 |
| *ReviewerVariance* | 0.51 | 0.37 | 0 | 4.00 |
| *RestReviewCnt* | 37.16 | 58.65 | 1 | 586 |
| *RestRating* | 3.87 | 0.48 | 1 | 5 |
| *RestVariance* | 0.54 | 0.39 | 0 | 4.00 |

## 5. Empirical Analyses and Results

### 5.1. Review Characteristics and Review Helpfulness/Unhelpfulness Scores

We first conducted a review-level analysis to examine how review characteristics affect helpfulness/unhelpfulness scores. Specifically, our analyses were based on the following regression specification:

$$
\begin{aligned}
DV = {} & \alpha_0 + \alpha_1 Rating + \alpha_2 Rating^2 + \alpha_3 Ln(Length) + \alpha_4 Ln(Photo) + \alpha_5 Topic_1 + \\
& \alpha_6 Topic_2 + \alpha_7 Topic_3 + \alpha_8 Sentiment + \alpha_9 Arousal + \alpha_{10} Readability + \\
& \alpha_{11} Ln(Rank) + \alpha_{12} Ln(View) + \alpha_{13} Ln(ReviewerReviewCnt) + \\
& \alpha_{14} ReviewerRating + \alpha_{15} ReviewerVariance + \alpha_{16} Ln(RestReviewCnt) + \\
& \alpha_{17} RestRating + \alpha_{18} RestVariance + \varepsilon \quad\quad (1)
\end{aligned}
$$

Since our dependent variables (*Helpfulness* and *Unhelpfulness*) were count data, we used a negative binomial regression for our analysis. We also added the quadratic term for *Rating* because studies suggested that the relationship between review extremity and helpfulness is nonlinear [10]. Additionally, we applied the natural log transformation to variables that were skewed in nature as $ln(1 + x)$, for which x is the variable of interest.

In Table 3, we report the results of our review-level analyses. The first column presents our regression results, for which the main dependent variable is *Helpfulness*. Our results are largely consistent with those of previous studies (e.g., [10]). The positive coefficient of *Rating* and the negative coefficient of *Rating²* indicate an inverted-U relationship between rating and helpfulness scores. In other words, reviews that had ratings either too high or too low were associated with lower helpfulness scores compared with reviews that had moderate ratings. In addition, longer reviews and reviews with more photos tended to receive higher helpfulness scores because they were perceived as more informative and of higher quality. The coefficient of *Ln(Rank)* was negatively significant, indicating that reviews on the top tended to receive higher helpfulness scores. Intuitively, reviews with more exposure received higher helpfulness scores.

**Table 3. Impact of Review Characteristics on Helpfulness/Unhelpfulness Scores**

| VARIABLES | (1) Helpfulness | (2) Unhelpfulness |
|---|---|---|
| $Rating$ | 0.108 | -0.802*** |
|  | (0.072) | (0.138) |
| $Rating^2$ | -0.021** | 0.101*** |
|  | (0.010) | (0.020) |
| $Ln(Length)$ | 0.188*** | 0.095 |
|  | (0.035) | (0.053) |
| $Ln(Photo)$ | 0.339*** | 0.107 |
|  | (0.052) | (0.060) |
| Topic1 | 0.267*** | 0.143 |
|  | (0.044) | (0.118) |
| Topic2 | 0.360*** | 0.144 |
|  | (0.039) | (0.091) |
| Topic3 | 0.259*** | -0.049 |
|  | (0.059) | (0.128) |
| Sentiment | -0.034 | 0.011 |
|  | (0.059) | (0.166) |
| Arousal | -0.079** | 0.089 |
|  | (0.039) | (0.125) |
| Readability | -0.012*** | 0.006 |
|  | (0.004) | (0.008) |
| $Ln(Rank)$ | -0.392*** | -0.186 |
|  | (0.085) | (0.392) |
| $Ln(View)$ | 0.376*** | 0.748*** |
|  | (0.023) | (0.039) |
| $Ln(ReviewerReviewCnt)$ | 0.304*** | -0.163*** |
|  | (0.023) | (0.033) |
| $ReviewerRating$ | -0.049 | -0.113 |
|  | (0.067) | (0.072) |
| $ReviewerVariance$ | -0.262*** | -0.101 |
|  | (0.072) | (0.095) |
| $Ln(RestReviewCnt)$ | 0.384*** | 0.113 |
|  | (0.081) | (0.364) |
| $RestRating$ | -0.104*** | 0.005 |
|  | (0.015) | (0.049) |
| $RestVariance$ | -0.146*** | 0.030 |
|  | (0.016) | (0.062) |
| Constant | -1.282*** | -3.792*** |
|  | (0.488) | (0.734) |
| Pseudo R-square | 0.129 | 0.066 |
| Observations | 157,285 | 157,285 |

Note: Standard errors in parentheses are robust and clustered by reviewers. *** $p < 0.01$, ** $p < 0.05$, * $p < 0.1$

In terms of textual characteristics, compared with reviews heavily focusing on Topic 4 (i.e., restaurant infrastructure), reviews that focused on food, drinks, and services (i.e., topics 1-3) tended to receive higher helpfulness scores. Also, the coefficient of text sentiment was negative but insignificant. The coefficient of text arousal meanwhile was negatively significant, indicating that reviews that contain high intensity of emotion tended to receive lower helpfulness scores. The GF index was also negatively associated with helpfulness scores. In other words, reviews with elusive text tend to attain lower helpfulness scores. In addition, reviews generated by experienced reviewers (i.e., those who wrote many reviews in the past) were more likely to obtain more helpfulness votes. In contrast, reviews generated by reviewers who provide a variety of ratings were less likely to receive helpfulness votes. Finally, reviews written for highly reviewed restaurants tended to obtain higher helpfulness scores, whereas reviews on restaurants with higher rating or larger rating variance tended to receive lower helpfulness scores.

In the second column, we report the results related to the primary interest of our study (i.e., where the main dependent variable is the *Unhelpfulness* score). Intuitively, if review readers cast unhelpfulness votes based on intrinsic review characteristics and restaurant characteristics, we would then expect to observe statistically significant relationships between independent variables and the dependent variable, with opposite signs compared to our results in Column (1). Interestingly, we observe this pattern only for the coefficients of *Rating* and *Rating$^2$*. The coefficients of all other variables reflecting reviews' intrinsic characteristics were statistically insignificant. The coefficient of *Ln(View)* was statistically significant, which indicated that reviews that attracted more readers tended to receive higher review unhelpfulness scores. This finding is consistent with intuition based on the effect of exposure. We also observe that reviews generated by experienced reviewers who wrote more reviews in the past tended to receive fewer unhelpfulness votes. This result demonstrates that review unhelpfulness scores may not be largely driven by the same intrinsic review characteristics that drive review helpfulness scores, as suggested by the meta goal-based choice model. We also formally test the differences between the coefficients of each independent variables for models (1) and (2). Using the Z-test for this comparison, we find that each coefficient is statistically different at p-value < 0.05. As a result, our Hypothesis 1 is supported. This finding should caution academic researchers who utilize review unhelpfulness votes as a proxy for identifying reviews with lower quality (e.g., deducting unhelpfulness scores from total scores to determine helpfulness degree or review quality) since review unhelpfulness scores do not appear to be driven by intrinsic review characteristics that are generally considered proxies for review quality (e.g., review length, review readability). To address potential concerns regarding the imbalance of helpfulness votes and unhelpfulness votes in our dataset, we perform two robustness tests in which we undersample reviews with helpfulness votes in one test and oversample reviews with unhelpfulness votes in another test. We report our results, which are qualitatively similar to our main results, in Tables A4 and A5 in the Appendix.

## 5.2 The Generation of Unhelpfulness Scores

As we discussed earlier, balance theory suggests that review unhelpfulness scores may be generated by reviewers whose reviews were voted unhelpful in the past, for such reviewers seek to rebalance their state of mind by casting unhelpfulness votes for other reviews. To test this hypothesis, we tracked all individuals who voted for 2016 reviews and who also contributed reviews themselves (to control for the influence of their review writing behavior). We identified 2,033 such voters and constructed voter-monthly-level panel data spanning the year 2016. We report summary statistics of this dataset in Table 4, and also report correlations among these variables in Table A3 in the Appendix.

**Table 4 Summary Statistics of Voter-level Data**

| Variable | Mean | SD | Min | Max |
|---|---|---|---|---|
| $HelpfulSent_{it}$ | 100.62 | 422.92 | 0 | 7,575 |
| $UnhelpfulSent_{it}$ | 0.24 | 3.29 | 0 | 232 |
| $Unhelpful\%_{it-1}$ | 0.01 | 0.05 | 0 | 1 |
| $Votes_{it-1}$ | 100.12 | 317.25 | 0 | 4,750 |
| $ReviewCount_{it}$ | 9.40 | 14.95 | 1 | 436 |
| $Rating_{it}$ | 3.83 | 0.60 | 1 | 5 |
| $Length_{it}$ | 569.23 | 648.87 | 0 | 18,948 |
| $Photo_{it}$ | 6.01 | 4.66 | 0 | 100 |
| $VoterReviewExperience_{it}$ | 164.49 | 240.74 | 1 | 2,789 |

We then perform a fixed effects panel data analysis with the following regression specification:

$$DV_{it} = \beta_0 + \beta_1 Unhelpful\%_{it-1} + \beta_2 \text{Ln}(Votes_{it-1}) + \beta_3 \text{Ln}(ReviewCount_{it}) +$$
$$\beta_4 Rating_{it} + \beta_5 \text{Ln}(Length_{it}) + \beta_6 \text{Ln}(Photo_{it}) +$$
$$\beta_7 \text{Ln}(VoterReviewExperience_{it-1}) + A_i + B_t + \varepsilon_{it} \qquad (2)$$

The dependent variables in this analysis were helpfulness and unhelpfulness votes sent by user $i$ in month $t$ ($HelpfulSent_{it}$ and $UnhelpfulSent_{it}$). We used the panel negative binomial regression model. The main independent variable of interest was the proportion of unhelpfulness votes among all votes received by the focal user in the previous month ($Unhelpful\%_{it-1}$). We also controlled for the impacts of total votes received in the previous month ($\text{Ln}(Votes_{it-1})$), volume ($\text{Ln}(ReviewCount_{it})$), average rating ($Rating_{it}$), average length ($\text{Ln}(Length_{it})$), and average number of photos ($\text{Ln}(Photo_{it})$) in reviews generated by the focal voter in the current month, as well as the total number of reviews generated by the focal voter since joining the platform ($\text{Ln}(VoterReviewExperience_{it})$) to capture the potential impact of a voter's evolving experience. In addition, we included voter fixed effects, $A_i$, and month fixed effects, $B_t$, in our model to account for voter invariant and time invariant effects. Nevertheless, as explained by Allison and Waterman [48] and Greene [49], since this model is not a traditional fixed-effects model, our results should be interpreted accordingly.

**Table 5 Voter-level Panel Data Regression**

| VARIABLES | (1) $HelpfulSent_{it}$ | (2) $UnhelpfulSent_{it}$ |
|---|---|---|
| $Unhelpful\%_{it-1}$ | 0.193 | 1.499*** |
|  | (0.261) | (0.548) |
| $Ln(Votes_{it-1})$ | 0.178*** | 0.018 |
|  | (0.012) | (0.036) |
| $Ln(ReviewCount_{it})$ | 0.614*** | 0.649*** |
|  | (0.015) | (0.050) |
| $Rating_{it}$ | 0.046* | -0.064 |
|  | (0.025) | (0.083) |
| $Ln(Length_{it})$ | 0.268*** | 0.349*** |
|  | (0.023) | (0.090) |
| $Ln(Photo_{it})$ | 0.246*** | 0.108 |
|  | (0.030) | (0.095) |
| $Ln(VoterReviewExperience_{it-1})$ | -0.158*** | -0.423*** |
|  | (0.016) | (0.066) |
| Voter Fixed Effects | Yes | Yes |
| Month Fixed Effects | Yes | Yes |
| Constant | -3.391*** | -2.093*** |
|  | (0.174) | (0.666) |
| Observations | 11,380 | 4,741 |

Note: Standard errors in parentheses are robust and clustered by reviewers. *** p < 0.01, ** p < 0.05, * p < 0.1. The number of users drops in this analysis because not every user writes reviews or casts votes in each time period.

We present the results of our voter-level panel data analysis in Table 5. Column (1) shows our results when the dependent variable is $Helpfulness_{it}$, and Column (2) shows our results when the dependent variable is $Unhelpfulness_{it}$. Although the proportion of unhelpfulness votes among total votes obtained by the focal user in the previous period did not affect helpfulness votes sent in the current period, it is nonetheless positively associated with unhelpfulness votes sent in the current period. In other words, following balance theory, users indeed tended to rebalance their state of mind after receiving unhelpfulness votes by voting other reviews as unhelpful, which provides support for Hypothesis 2.

We acknowledge that endogeneity issues in our analysis might exist. For example, although we controlled for the impact of voters' evolving experience, there might be other factors that impact their review evaluating (i.e., casting vote) and generating behavior, which then influences helpfulness votes received at the same time. As a result, we may have difficulty correctly identifying the impact of receiving unhelpfulness votes. Therefore, we employed a quasi-experimental research design that leveraged the dynamic propensity score matching (dPSM) technique to account for this issue [50, 51]. Specifically, we considered users who received helpfulness votes/unhelpfulness votes in a given period as the treatment group and considered users who did not receive a vote in that period as the control group. Receipt of helpfulness votes was considered a different treatment condition than receipt of unhelpfulness votes. We matched each treated user to a control user based on their propensity score of receiving votes in a given period. We then compared the vote-casting behavior of treated users to that of matched control users in that period to assess the impact of receiving helpfulness or unhelpfulness votes. We repeated this matching and assessing technique for each month in 2016.

We incorporated four sets of matching covariates. First, we captured users' review-generating behavior, including the accumulative volume (*ReviewCntAcc*), average star rating (*RatingAcc*), and average length (*LengthAcc*) of reviews generated by the focal user. Second, we captured users' vote-receiving and -sending behavior using the number of helpfulness/unhelpfulness votes that users received (*HelpfulnessAcc, UnhelpfulnessAcc*) and sent (*HelpfulSentAcc,*

*UnhelpfulSentAcc*). Third, we included users' gender and the number of months since they joined the platform (*Gender*, *Tenure*). Lastly, we captured users' restaurant preferences by averaging the characteristics of restaurants that the focal user reviewed, including the volume, rating, and rating variance of reviews that the restaurant received (*RestReviewCntAcc, RestRatingAcc, RestVarAcc*). In Table 6, we show all matching covariates and their measures.

**Table 6 Measures of Matching Covariates**

| Variable | Measure |
|---|---|
| Review-generating | |
| *ReviewCntAcc* | Number of reviews generated by a focal user $i$ until month $t–1$ |
| *RatingAcc* | Average star rating of reviews generated by a focal user $i$ until month $t–1$ |
| *LengthAcc* | Average word counts of reviews generated by a focal user $i$ until month $t–1$ |
| Vote-receiving and -sending | |
| *HelpfulnessAcc* | Helpfulness votes received by a focal user $i$ until month $t–1$ |
| *UnhelpfulnessAcc* | Unhelpfulness votes received by a focal user $i$ until month $t–1$ |
| *HelpfulSentAcc* | Helpfulness votes sent by a focal user $i$ until month $t–1$ |
| *UnhelpfulSentAcc* | Unhelpfulness votes sent by a focal user $i$ until month $t–1$ |
| Users' demographic | |
| *Gender* | Gender of user $i$ |
| *Tenure* | Number of months since a focal user joined the platform in month $t$ |
| Restaurant choice | |
| *RestReviewCntAcc* | Average number of reviews received by restaurants that were reviewed by a focal user $i$ until month $t–1$ |
| *RestRatingAcc* | Average star rating received by restaurants that were reviewed by a focal user $i$ until month $t–1$ |
| *RestVarAcc* | Rating variance of reviews received by restaurants that were reviewed by a focal user $i$ until month $t–1$ |

We then conducted PSM using a Probit model. We employed the single nearest neighbor-matching method to obtain a one-to-one matched control user for each of the treated users. We specified the common support option and a caliper of 0.05. After matching, we conducted paired t-tests on each of the covariates to determine whether there was a significant difference between the treatment and control groups. As we observed in Table 7, all covariates were balanced after matching. Specifically, all matching covariates had a $p>0.10$ difference between the treated and control groups for *HelpfulSent* and *UnhelpfulSent*. This test confirmed that our matching attempt successfully generated a control group similar to the treated group. Figure 1 demonstrates the pairwise mean difference in the number of helpfulness votes cast between treated and control users across the 12 months with 95% confidence intervals.

**Table 7 Balance Test After Matching (1st Month)**

| Variable | *HelpfulSent* | | | | *UnhelpfulSent* | | | |
| --- | --- | --- | --- | --- | --- | --- | --- | --- |
| | Mean (T) | Mean (C) | T-Value | P-Value | Mean (T) | Mean (C) | T-Value | P-Value |
| *Ln(ReviewCntAcc)* | 2.48 | 2.50 | −0.404 | 0.686 | 4.31 | 4.35 | −0.316 | 0.753 |
| *RatingAcc* | 3.94 | 3.95 | −0.451 | 0.652 | 3.77 | 3.79 | −0.505 | 0.614 |
| *Ln(LengthAcc)* | 5.41 | 5.35 | 1.172 | 0.242 | 5.99 | 6.02 | −0.349 | 0.728 |
| *Ln(HelpfulnessAcc)* | 3.31 | 3.37 | −0.916 | 0.340 | 5.70 | 5.73 | −0.104 | 0.917 |
| *Ln(UnhelpfulnessAcc)* | 0.27 | 0.29 | −1.327 | 0.185 | 1.29 | 1.30 | −0.061 | 0.951 |
| *Ln(HelpfulnessSentAcc)* | 2.32 | 2.40 | −1.048 | 0.295 | 4.61 | 4.68 | −0.232 | 0.816 |
| *Ln(UnhelpfulnessSentAcc)* | 0.24 | 0.25 | −0.266 | 0.790 | 0.85 | 0.80 | −0.511 | 0.610 |
| *Ln(Tenure)* | 3.14 | 3.19 | −1.381 | 0.168 | 3.30 | 3.22 | 0.890 | 0.374 |
| *Gender* | 0.64 | 0.62 | 0.876 | 0.381 | 0.67 | 0.67 | 0.013 | 0.989 |
| *Ln(RestReviewCntAcc)* | 2.90 | 2.79 | 1.072 | 0.284 | 2.98 | 2.84 | 1.434 | 0.152 |
| *RestRatingAcc* | 3.92 | 3.92 | −0.261 | 0.794 | 3.85 | 3.85 | 0.058 | 0.954 |
| *RestVarAcc* | 0.56 | 0.55 | 0.355 | 0.723 | 0.54 | 0.55 | −0.097 | 0.923 |

Note: T and C indicate the treatment group and the control group, respectively.

**Figure 1 Effect of Helpfulness Votes Received on Helpfulness Votes Sent**

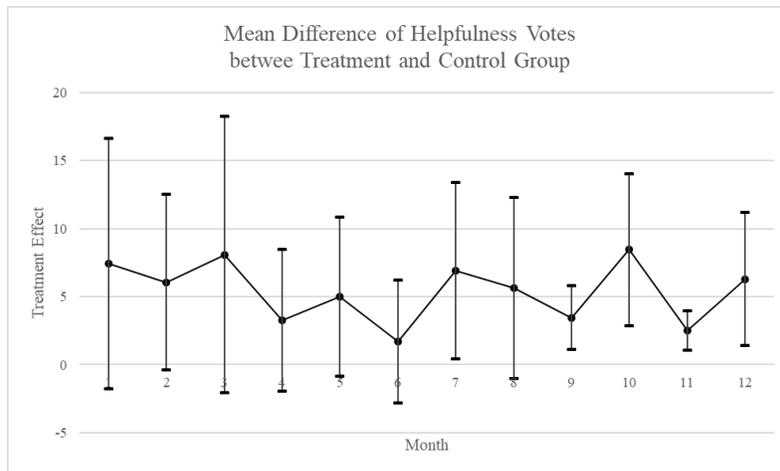

Although the mean difference of the two groups across the entire study period was 5.36 (p<0.001), it was not statistically significant for any single month. In Figure 2, we show the effect of receiving unhelpfulness votes on unhelpfulness vote sending. The mean difference across the entire study period was 0.40 (p<0.001); it was significant at the 0.05 level in each month of 2016. Consistent with our result from our panel data analysis, we found that when users received helpfulness votes, they did not reciprocate by casting helpfulness votes to others. However, when users received unhelpfulness votes, they were more likely to vote for other reviews as unhelpful.

**Figure 2 Effect of Unhelpfulness Votes Received on Unhelpfulness Votes Sent**

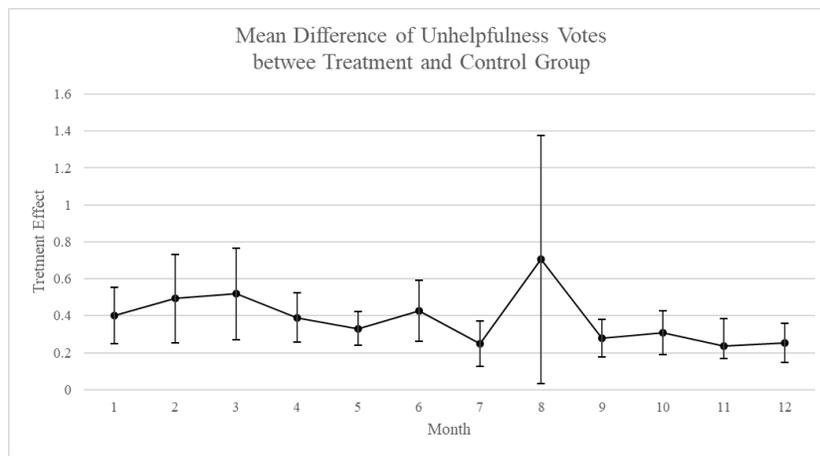

## 5.3 Post-hoc Analysis of Helpfulness Voters vs Unhelpfulness Voters

In this subsection, we answer our second research question by comparing how unhelpfulness voters differ from helpfulness voters with respect to platform engagement. We define helpfulness voters as users who only cast review helpfulness votes, define unhelpfulness voters as those who only cast review unhelpfulness votes, and define mixed voters as those who cast both types of votes. There were 22,865 helpfulness voters, 2,857 unhelpfulness voters, and 1,491 mixed voters in our dataset, respectively.

We performed an ANOVA test on a set of variables that reflected users' engagement with the review platform, including the review volume (*ReviewCount*), the average content length (*Length*), and the average photos per review (*Photo*) generated by these users before they cast their first vote in 2016. We present our ANOVA results and post-hoc pairwise comparison tests for means in Tables 8 and 9.

We found that mixed voters exhibited the highest level of engagement, followed by helpfulness voters. Unhelpfulness voters engaged the least with the platform, writing 4.70 reviews on average from inception. This number was lower than helpfulness voters' 7.01 reviews and far lower than mixed voters' 55.42 reviews. The average review length was also significantly different. The average length of reviews written by unhelpfulness voters was 263.14, whereas that of helpfulness voters and mixed voters was 264.75 and 388.20, respectively. With respect to photo use in reviews, unhelpfulness voters averaged only 2.32 photos, whereas helpfulness voters and mixed voters attached 3.10 photos and 3.80 photos in their reviews on average. In sum, pure unhelpfulness voters behave significantly differently from helpfulness voters. The former appear to be less engaged with the platform.

These results indicate that mixed voters are likely to read reviews more thoroughly, so they may make a balanced judgment. Their willingness to engage in both types of voting suggests a higher level of involvement with the platform. Helpfulness voters might prefer to highlight useful information rather than focus on negating what they do not find useful. This selective engagement suggests that helpfulness voters are involved with content, but that their interaction is more limited compared with that of mixed voters. Unhelpfulness voters demonstrate the lowest level of engagement, which might reflect a more critical or selective approach to content.

Specifically, this behavior suggests more passive or limited interaction with a platform's content.

Table 8. ANOVA Result of Voter Engagement

|  | Sum of Squares | d.f. | Mean Square | F-Ratio | Significance |
|---|---|---|---|---|---|
| *ReviewCount* | | | | | |
| Between groups | 3,345,897.53 | 2 | 1,672,948.77 | 670.52 | 0.000 |
| Within group | 67,889,053.6 | 27,210 | 2,495.00 | | |
| Total | 71,234,951.1 | 27,212 | 2617.78 | | |
| *Length* | | | | | |
| Between groups | 15,598,643 | 2 | 7,799,321.49 | 134.75 | 0.000 |
| Within group | 715,639,243 | 12,364 | 57,880.88 | | |
| Total | 731,237,886 | 12,366 | 59,132.94 | | |
| *Photo* | | | | | |
| Between groups | 1,418.23 | 2 | 709.17 | 50.15 | 0.000 |
| Within group | 201,257.57 | 14,232 | 14.14 | | |
| Total | 202,675.81 | 14,234 | 14.24 | | |

Table 9. ANOVA Post-hoc Pairwise Comparison Tests for the Means
(p-value adjusted by Bonferroni correction)

| Comparison | Mean Difference | Significance |
|---|---|---|
| *ReviewCount* | | |
| Hvoter vs. Uvoter | –2.37 | 0.050 |
| Hvoter vs. Mvoter | 48.36 | 0.000 |
| Uvoter vs. Mvoter | 50.73 | 0.000 |
| *Length* | | |
| Hvoter vs. Uvoter | –1.61 | 1.000 |
| Hvoter vs. Mvoter | 125.06 | 0.000 |
| Uvoter vs. Mvoter | 123.44 | 0.000 |
| *Photo* | | |
| Hvoter vs. Uvoter | –0.78 | 0.000 |
| Hvoter vs. Mvoter | 0.70 | 0.000 |
| Uvoter vs. Mvoter | 1.48 | 0.000 |

Note: Hvoter, Uvoter, and Mvoter represent helpfulness voters, unhelpfulness voters, and mixed voters, respectively.

## 6. Discussion and Conclusion

Review platforms have increasingly incorporated review evaluation systems to assist consumers'

decision-making, improve their satisfaction, and enhance content quality. Although many platforms adopt review evaluation systems that allow users to vote for both helpful and unhelpful reviews, most review evaluation studies focus only on reviewing helpfulness scores. As a result, our knowledge of the nature of review unhelpfulness scores is limited. Nevertheless, studies that analyze review helpfulness and unhelpfulness scores implicitly assume that both scores are not only related to reviews' intrinsic characteristics, but that both scores can also be consistently used to identify high-quality and low-quality reviews, respectively. The primary purpose of our study was to examine such conventional wisdom by paying specific attention to factors that influence review unhelpfulness scores. Specifically, we employed meta goal-based choice model to investigate whether review unhelpfulness scores are driven by intrinsic review characteristics that influence review helpfulness scores. We also drew upon balance theory to examine factors that drive the generation of review helpfulness scores. Lastly, we uncovered systematic differences between users who cast helpfulness votes for reviews versus those who cast unhelpfulness votes for reviews.

Using a unique dataset that we obtained through a collaboration with a large restaurant review platform in Asia, our first analysis demonstrated that, unlike the case with review helpfulness scores, review unhelpfulness scores do not appear to be driven by intrinsic review characteristics—with the exception of review ratings, which coincides with the logic of meta goal-based choice model. In addition, as posited by balance theory, users who receive unhelpfulness votes tend to vote for other reviews as unhelpful. Furthermore, unhelpfulness voters are significantly different from helpfulness voters with respect to platform engagement. Our results empirically demonstrate that, unlike the case with review helpfulness scores, review unhelpfulness scores may not be correlated with intrinsic review characteristics that proxy for review quality. As such, we caution researchers and practitioners who use review unhelpfulness scores to infer review quality.

Our study yields important implications for research and practice. From a research perspective, our work is among the first to comprehensively identify factors that influence review unhelpfulness scores. As such, future studies can use our empirical findings to further expand research horizons in this important area. In addition, we connect our empirical findings with meta goal-based choice model, which provides a theoretical foundation for future studies that seek to identify underlying mechanisms that drive the behavior of unhelpfulness voters. Lastly, our paper provides empirical evidence that cautions against the use of unhelpfulness votes to identify reviews with low quality; this evidence, which is lacking in the literature, has significant implications for online review research.

Our findings also offer several important and actionable insights for practitioners, especially for managers of platforms that utilize online reviews. First, our empirical results can help platform managers develop policies and interventions related to incentives for review contributions. In particular, for performance-contingent incentive policies, we show that while review helpfulness scores can be used as one of the selection criteria for reviews with high quality, review unhelpfulness scores should not be used to identify reviews with low quality. In addition, our results show that review ratings significantly influence unhelpfulness votes. As such, platform managers who prefer to use review unhelpfulness scores to develop policies and interventions should account for ratings of reviews that receive unhelpfulness votes. Lastly, our findings regarding the differences among helpfulness voters, unhelpfulness voters, and mixed voters can also be used to monitor users and develop incentive policies that encourage and retain platform users.

Our study is not without limitations, which also provide avenues for future research. First, our study relies on observation data; even though our empirical results are in line with theoretical predictions from our hypotheses, we are still unable to pinpoint the exact underlying mechanisms that drive users to cast unhelpfulness votes. Future qualitative studies that can access users' state of mind (e.g., user interviews, surveys) could further identify such mechanisms. Second, as our study uses data from an Asian online review platform, we note that cultural factors may play a role in helpfulness and unhelpfulness vote casting. Therefore, generalizing our results to other markets must be performed cautiously. Thus, future research on different markets might explore the impact of cultural factors in voting patterns. Finally, future research could examine whether other online platforms that utilize peer evaluation systems (e.g., online discussion platforms, content-sharing platforms) experience similar vote-casting patterns that we observed in our study.

# APPENDIX

**Table A1 Studies Using the Ratio of Helpfulness Votes to Total Votes to Measure Review Helpfulness/Quality**

| No | Study |
|---|---|
| 1 | Chen, P.-Y., Dhanasobhon, S., and Smith, M. D. 2008. "All Reviews Are Not Created Equal: The Disaggregate Impact of Reviews and Reviewers at Amazon.com." Working paper. |
| 2 | Chua, A. Y., and Banerjee, S. 2015. "Understanding Review Helpfulness as a Function of Reviewer Reputation, Review Rating, and Review Depth," *Journal of the Association for Information Science and Technology*, (66:2), pp. 354–362. |
| 3 | Forman, C., Ghose A., and Wiesenfeld, B. 2008. "Examining the Relationship Between Reviews and Sales: The Role of Reviewer Identity Disclosure in Electronic Markets," *Information Systems Research* (19:3), pp. 291–313. |
| 4 | Kim, S. J., Maslowska, E., and Tamaddoni, A. 2019. "The Paradox of (dis) Trust in Sponsorship Disclosure: The Characteristics and Effects of Sponsored Online Consumer Reviews," *Decision Support Systems*, 116, pp. 114–124. |
| 5 | Korfiatis, N., García-Bariocanal, E., and Sánchez-Alonso, S. 2012. "Evaluating Content Quality and Helpfulness of Online Product Reviews: The Interplay of Review Helpfulness vs. Review Content," *Electronic Commerce Research and Applications*, (11:3), pp. 205–217. |
| 6 | Kuan, K. K., Hui, K. L., Prasarnphanich, P., and Lai, H. Y. 2015. "What Makes a Review Voted? An Empirical Investigation of Review Voting in Online Review Systems," *Journal of the Association for Information Systems*, (16:1), pp. 48–71. |
| 7 | Lei, Z., Yin, D., and Zhang, H. 2020. "Focus Within or On Others: The Impact of Reviewers' Attentional Focus on Review Helpfulness," *Information Systems Research*, (32:3), pp. 801-819. |
| 8 | Mudambi, S. M., and Schuff, D. 2010. "What Makes a Helpful Review? A Study of Customer Reviews on Amazon.com," *MIS Quarterly* (34:1), pp. 185–200. |
| 9 | Wang, Y., Goes, P., Wei, Z., and Zeng, D. 2019. "Production of Online Word-of-Mouth: Peer Effects and the Moderation of User Characteristics," *Production and Operations Management* (28:7), pp. 1621–1640. |
| 10 | Yin, D., Bond, S. D., and Zhang, H. 2014. "Anxious or Angry? Effects of Discrete Emotions on the Perceived Helpfulness of Online Reviews," *MIS Quarterly* (38:2), pp. 539–560. |
| 11 | Yin, D., Bond, S. D., and Zhang, H. 2017. "Keep Your Cool or Let It Out: Nonlinear Effects of Expressed Arousal on Perceptions of Consumer Reviews." *Journal of Marketing Research* (54:3), pp. 447–463. |
| 12 | Zhang, R., and Tran, T. 2010. "Helpful or Unhelpful: A Linear Approach for Ranking Product Reviews," *Journal of Electronic Commerce Research* (11:3), pp. 220–230. |

**Table A2 Correlations for Variables Used in the Review-level Analysis**

|  | 1 | 2 | 3 | 4 | 5 | 6 | 7 | 8 | 9 | 10 | 11 | 12 | 13 | 14 | 15 | 16 | 17 | 18 | 19 |
|---|---|---|---|---|---|---|---|---|---|---|---|---|---|---|---|---|---|---|---|
| *Helpfulness* | 1.000 | | | | | | | | | | | | | | | | | | |
| *Unhelpfulness* | 0.108 | 1.000 | | | | | | | | | | | | | | | | | |
|  | (0.000) | | | | | | | | | | | | | | | | | | |
| *Rating* | -0.101 | -0.011 | 1.000 | | | | | | | | | | | | | | | | |
|  | (0.000) | (0.000) | | | | | | | | | | | | | | | | | |
| *Length* | 0.441 | 0.062 | -0.056 | 1.000 | | | | | | | | | | | | | | | |
|  | (0.000) | (0.000) | (0.000) | | | | | | | | | | | | | | | | |
| *Photo* | 0.375 | 0.044 | 0.012 | 0.510 | 1.000 | | | | | | | | | | | | | | |
|  | (0.000) | (0.000) | (0.000) | (0.000) | | | | | | | | | | | | | | | |
| *Topic1* | 0.012 | -0.006 | 0.016 | 0.018 | -0.001 | 1.000 | | | | | | | | | | | | | |
|  | (0.000) | 0.007 | (0.000) | (0.000) | 0.754 | | | | | | | | | | | | | | |
| *Topic2* | 0.072 | 0.003 | 0.044 | 0.015 | 0.022 | -0.431 | 1.000 | | | | | | | | | | | | |
|  | (0.000) | 0.125 | (0.000) | (0.000) | (0.000) | (0.000) | (0.000) | | | | | | | | | | | | |
| *Topic3* | 0.055 | -0.005 | 0.001 | 0.013 | 0.044 | -0.171 | -0.288 | 1.000 | | | | | | | | | | | |
|  | (0.000) | 0.015 | 0.826 | (0.000) | (0.000) | (0.000) | (0.000) | | | | | | | | | | | | |
| *Sentiment* | -0.104 | -0.014 | 0.348 | -0.143 | -0.059 | -0.035 | 0.055 | -0.103 | 1.000 | | | | | | | | | | |
|  | (0.000) | (0.000) | (0.000) | (0.000) | (0.000) | (0.000) | (0.000) | (0.000) | | | | | | | | | | | |
| *Arousal* | -0.066 | 0.002 | 0.069 | -0.084 | -0.055 | -0.059 | 0.100 | -0.123 | 0.197 | 1.000 | | | | | | | | | |
|  | (0.000) | 0.471 | (0.000) | (0.000) | (0.000) | (0.000) | (0.000) | (0.000) | (0.000) | | | | | | | | | | |
| *Readability* | -0.029 | -0.001 | 0.044 | -0.018 | 0.003 | -0.023 | -0.021 | 0.077 | 0.109 | 0.025 | 1.000 | | | | | | | | |
|  | (0.000) | 0.732 | (0.000) | (0.000) | 0.176 | (0.000) | (0.000) | (0.000) | (0.000) | (0.000) | | | | | | | | | |
| *Rank* | -0.036 | -0.002 | 0.052 | -0.013 | -0.027 | -0.063 | 0.017 | -0.106 | 0.058 | 0.020 | -0.013 | 1.000 | | | | | | | |
|  | (0.000) | 0.303 | (0.000) | (0.000) | (0.000) | (0.000) | (0.000) | (0.000) | (0.000) | (0.000) | (0.000) | | | | | | | | |
| *View* | 0.442 | 0.135 | -0.004 | 0.391 | 0.301 | -0.017 | 0.019 | 0.002 | -0.055 | -0.032 | -0.011 | 0.046 | 1.000 | | | | | | |
|  | (0.000) | (0.000) | 0.090 | (0.000) | (0.000) | (0.000) | (0.000) | 0.370 | (0.000) | (0.000) | (0.000) | (0.000) | | | | | | | |
| *ReviewerReviewCnt* | 0.411 | 0.011 | -0.160 | 0.232 | 0.248 | 0.061 | 0.040 | 0.053 | -0.102 | -0.067 | 0.000 | -0.079 | 0.197 | 1.000 | | | | | |
|  | (0.000) | (0.000) | (0.000) | (0.000) | (0.000) | (0.000) | (0.000) | (0.000) | (0.000) | (0.000) | 0.985 | (0.000) | (0.000) | | | | | | |
| *ReviwerRating* | -0.171 | -0.012 | 0.340 | -0.111 | -0.085 | -0.028 | 0.018 | -0.007 | 0.137 | 0.053 | 0.021 | 0.040 | -0.068 | -0.251 | 1.000 | | | | |
|  | (0.000) | (0.000) | (0.000) | (0.000) | (0.000) | (0.000) | (0.000) | 0.004 | (0.000) | (0.000) | (0.000) | (0.000) | (0.000) | (0.000) | | | | | |
| *ReviewerVariance* | -0.015 | -0.001 | -0.083 | 0.017 | -0.025 | 0.000 | -0.040 | -0.041 | -0.052 | 0.000 | -0.017 | 0.002 | 0.001 | -0.033 | -0.272 | 1.000 | | | |
|  | (0.000) | 0.594 | (0.000) | (0.000) | (0.000) | 0.846 | (0.000) | (0.000) | (0.000) | 0.988 | (0.000) | 0.333 | 0.564 | (0.000) | (0.000) | | | | |
| *RestReviewCnt* | -0.036 | -0.002 | 0.052 | -0.013 | -0.027 | -0.063 | 0.017 | -0.106 | 0.058 | 0.020 | -0.013 | 1.000 | 0.046 | -0.079 | 0.040 | 0.002 | 1.000 | | |
|  | (0.000) | 0.303 | (0.000) | (0.000) | (0.000) | (0.000) | (0.000) | (0.000) | (0.000) | (0.000) | (0.000) | (0.000) | (0.000) | (0.000) | (0.000) | 0.333 | | | |
| *RestRating* | -0.030 | 0.010 | 0.301 | -0.002 | 0.009 | -0.042 | 0.059 | -0.006 | 0.140 | 0.052 | 0.019 | 0.076 | 0.044 | -0.084 | 0.097 | -0.026 | 0.076 | 1.000 | |
|  | (0.000) | (0.000) | (0.000) | 0.329 | (0.000) | (0.000) | (0.000) | 0.005 | (0.000) | (0.000) | (0.000) | (0.000) | (0.000) | (0.000) | (0.000) | (0.000) | (0.000) | | |
| *RestVariance* | -0.030 | 0.003 | -0.105 | 0.009 | 0.006 | -0.016 | -0.114 | -0.038 | -0.035 | -0.001 | -0.005 | 0.201 | 0.024 | -0.056 | -0.004 | 0.028 | 0.201 | -0.301 | 1.000 |
|  | (0.000) | 0.135 | (0.000) | (0.000) | 0.011 | (0.000) | (0.000) | (0.000) | (0.000) | 0.605 | 0.048 | (0.000) | (0.000) | (0.000) | 0.150 | (0.000) | (0.000) | (0.000) | |

**Table A3 Correlations for Variables Used in the Voter-level Analysis**

|  | 1 | 2 | 3 | 4 | 5 | 6 | 7 | 8 | 9 |
|---|---|---|---|---|---|---|---|---|---|
| *HelpfulSent* | 1.000 | | | | | | | | |
| *UnhelpfulSent* | 0.124 | 1.000 | | | | | | | |
|  | (0.000) | | | | | | | | |
| *Unhelpful%* | -0.033 | 0.024 | 1.000 | | | | | | |
|  | (0.001) | (0.013) | | | | | | | |
| *Votes* | 0.791 | 0.156 | -0.040 | 1.000 | | | | | |
|  | (0.000) | (0.000) | (0.000) | | | | | | |
| *ReviewCount* | (0.348) | (0.132) | -0.041 | 0.568 | 1.000 | | | | |
|  | (0.000) | (0.000) | (0.000) | (0.000) | | | | | |
| *Rating* | -0.089 | -0.019 | -0.029 | -0.105 | -0.124 | 1.000 | | | |
|  | (0.000) | (0.038) | 0.003 | (0.000) | (0.000) | | | | |
| *Length* | 0.201 | 0.057 | -0.013 | 0.206 | 0.050 | -0.018 | 1.000 | | |
|  | (0.000) | (0.000) | (0.189) | (0.000) | (0.000) | (0.050) | | | |
| *Photo* | 0.193 | 0.051 | -0.023 | 0.238 | 0.124 | 0.054 | 0.559 | | |
|  | (0.000) | (0.000) | 0.015 | (0.000) | (0.000) | (0.000) | (0.000) | | |
| *VoterReviewExperience* | 0.339 | 0.091 | -0.045 | 0.531 | 0.521 | -0.151 | 0.215 | 0.279 | |
|  | (0.000) | (0.000) | (0.000) | (0.000) | (0.000) | (0.000) | (0.000) | (0.000) | 1.000 |

**Table A4 Results of Review-level Analysis Using Undersampling**

| VARIABLES | Helpfulness | Unhelpfulness |
|---|---|---|
| $Rating$ | 0.128 | -0.344*** |
|  | (0.117) | (0.097) |
| $Rating^2$ | -0.021 | 0.043*** |
|  | (0.017) | (0.014) |
| $Ln(Length)$ | 0.225*** | 0.068* |
|  | (0.050) | (0.039) |
| $Ln(Photo)$ | 0.109* | -0.040 |
|  | (0.056) | (0.035) |
| Topic1 | 0.263*** | -0.015 |
|  | (0.084) | (0.085) |
| Topic2 | 0.267*** | 0.051 |
|  | (0.072) | (0.064) |
| Topic3 | 0.252** | -0.103 |
|  | (0.109) | (0.103) |
| Sentiment | 0.194 | -0.055 |
|  | (0.144) | (0.128) |
| Arousal | -0.095 | 0.119 |
|  | (0.091) | (0.088) |
| Readability | -0.008 | 0.003 |
|  | (0.006) | (0.006) |
| $Ln(Rank)$ | -0.401 | -0.293 |
|  | (0.308) | (0.307) |
| $Ln(View)$ | 0.316*** | 0.295*** |
|  | (0.028) | (0.025) |
| $Ln(ReviewerReviewCnt)$ | 0.335*** | -0.080*** |
|  | (0.028) | (0.017) |
| $ReviewerRating$ | -0.010 | -0.071* |
|  | (0.080) | (0.041) |
| $ReviewerVariance$ | -0.264** | -0.066 |
|  | (0.107) | (0.062) |
| $Ln(RestReviewCnt)$ | 0.370 | 0.241 |
|  | (0.285) | (0.283) |
| $RestRating$ | -0.126*** | 0.033 |
|  | (0.039) | (0.040) |
| $RestVariance$ | -0.139*** | 0.044 |
|  | (0.053) | (0.050) |
| Constant | -0.829 | -0.793 |
|  | (0.678) | (0.496) |
| Pseudo $R^2$ | 0.144 | 0.037 |
| Observations | 3,895 | 3,895 |

Note: Standard errors in parentheses are robust and clustered by reviewers. *** p < 0.01, ** p < 0.05, * p < 0.1

**Table A5 Results of Review-level Analysis Using Oversampling**

| VARIABLES | Helpfulness | Unhelpfulness |
|---|---|---|
| $Rating$ | -0.004** | 0.005 |
|  | (0.002) | (0.003) |
| $Rating^2$ | -0.001*** | 0.0005 |
|  | (0.0002) | (0.0005) |
| $Ln(Length)$ | 0.035*** | 0.007* |
|  | (0.002) | (0.004) |
| $Ln(Photo)$ | 0.031*** | 0.012 |
|  | (0.002) | (0.013) |
| Topic1 | 0.002 | -0.013 |
|  | (0.005) | (0.010) |
| Topic2 | 0.023*** | -0.018 |
|  | (0.004) | (0.008) |
| Topic3 | 0.022*** | 0.015 |
|  | (0.006) | (0.011) |
| Sentiment | -0.040*** | 0.008 |
|  | (0.009) | (0.016) |
| Arousal | -0.032*** | 0.015 |
|  | (0.008) | (0.015) |
| Readability | -0.0008* | -0.001* |
|  | (0.0004) | (0.001) |
| $Ln(Rank)$ | -0.001 | -0.002 |
|  | (0.001) | (0.002) |
| $Ln(View)$ | 0.030*** | 0.012*** |
|  | (0.002) | (0.003) |
| $Ln(ReviewerReviewCnt)$ | 0.014*** | 0.002* |
|  | (0.001) | (0.001) |
| $ReviewerRating$ | -0.027*** | 0.005 |
|  | (0.003) | (0.006) |
| $ReviewerVariance$ | -0.001 | 0.002 |
|  | (0.005) | (0.009) |
| $Ln(RestReviewCnt)$ | 0.0008 | 0.002 |
|  | (0.001) | (0.002) |
| $RestRating$ | -0.0002 | 0.005 |
|  | (0.003) | (0.006) |
| $RestVariance$ | -0.008* | -0.015* |
|  | (0.004) | (0.008) |
| Constant | 2.450*** | -2.379*** |
|  | (0.040) | (0.074) |
| Pseudo $R^2$ | 0.0014 | 0.0005 |
| Observations | 182,369 | 182,369 |

Note: Standard errors in parentheses are robust and clustered by reviewers. *** p < 0.01, ** p < 0.05, * p < 0.1